\documentclass[10pt]{article}
\usepackage{graphicx}

\usepackage{lineno}

\def\Title#1{\begin{center} {\Large #1 } \end{center}}
\def\Author#1{\begin{center}{ \sc #1} \end{center}}
\def\Address#1{\begin{center}{ \it #1} \end{center}}

\def\mtop{m_{\rm{top}}}

\def\mtoppole{m_{\rm{top}}^{\rm{pole}}}

\def\rtt{R_{\rm{3/2}}}

\def\ttbar{t\bar{t}}
\def\pp{pp}
\def\mlb{m_{\ell b}^{\rm{reco}}}
\def\Et{E_{\rm{T}}}
\def\pt{p_{\rm{T}}}
\def\met{E_{\rm{T}}^{\rm{miss}}}
\def\sigttbar{\sigma_{\ttbar}}

\def\GeV{\rm{GeV}}
\def\TeV{\rm{TeV}}
\def\pb{\rm{pb}}
\def\fbm1{\rm fb^{-1}}
\def\stat{\rm stat.}
\def\syst{\rm syst.}
\newcommand\pubblock{\rightline{\begin{tabular}{l} Proceedings of the Fifth Annual LHCP\\ \pubnumber\\
         \pubdate  \end{tabular}}}

\newenvironment{Abstract}{\begin{quotation} \begin{center} 
             \large ABSTRACT \end{center}\bigskip 
      \begin{center}\begin{large}}{\end{large}\end{center} \end{quotation}}

\newenvironment{Presented}{\begin{quotation} \begin{center} 
             PRESENTED AT\end{center}\bigskip 
      \begin{center}\begin{large}}{\end{large}\end{center} \end{quotation}}

\def\beq{\begin{equation}}
\def\eeq#1{\label{#1}\end{equation}}
\def\eeqn{\end{equation}}

\def\beqa{\begin{eqnarray}}
\def\eeqa#1{\label{#1}\end{eqnarray}}
\def\eeqan{\end{eqnarray}}

\let\bar=\overbar

\def\Dslash{\not{\hbox{\kern-4pt $D$}}}
\def\dslash{\not{\hbox{\kern-2pt $\del$}}}

\def\msb{{\bar{\ssstyle M \kern -1pt S}}}

\usepackage{color}

\newcommand\pubnumber{ATL-PHYS-PROC-2017-210}

\newcommand\pubdate{\today}

\def\affiliation{
On behalf of the ATLAS Collaboration \\
Max-Planck-Institute f\"{u}r Physik\\
F\"{o}hringer Ring 6, 80805 Muenchen, Germany }

\begin{document}


\large
\begin{titlepage}
\pubblock

\vfill
\Title{Top-quark mass and top-quark pole mass measurements with the ATLAS detector }
\vfill

\Author{ Teresa Barillari  }
\Address{\affiliation}
\vfill
\begin{Abstract}

Results of top-quark mass measurements in the di-lepton and 
the all-jets top-antitop decay channels with the ATLAS detector 
are presented. The measurements are obtained using proton--proton 
collisions at a centre-of-mass energy $\sqrt{s} = 8\,\TeV$ at 
the CERN Large Hadron Collider.
The data set used corresponds to an integrated luminosity of 
$20.2\,\fbm1$. The top-quark mass in the di-lepton channel is measured 
to be $172.99\,\pm\,0.41\,(\stat)\,\pm\,0.74\,(\syst)\,{\GeV}$. 
In the all-jets analysis the top-quark mass is measured to be 
$173.72 \,\pm\,0.55\,(\stat)\,\pm\,1.01\,(\syst)\,\GeV$.
In addition, the top-quark pole mass is determined from 
inclusive  cross-section measurements in the top-antitop di-lepton 
decay channel with the ATLAS detector. The measurements are obtained 
using data at $\sqrt{s} = 7\,\TeV$ and $\sqrt{s} =8\,\TeV$ corresponding to 
an integrated luminosity of $4.6\,\fbm1$ and $20.2\,\fbm1$ respectively.
The top-quark pole mass is measured to be $172.9^{+2.5}_{-2.6}\,\GeV$.

\end{Abstract}
\vfill

\begin{Presented}
The Fifth Annual Conference\\
 on Large Hadron Collider Physics \\
Shanghai Jiao Tong University, Shanghai, China\\ 
May 15-20, 2017
\end{Presented}
\vfill
\end{titlepage}
\def\thefootnote{\fnsymbol{footnote}}
\setcounter{footnote}{0}
%

\normalsize 


\section{Introduction}
\label{sec:intro}

Due to the higher centre-of-mass energy, top quark 
production at the proton--proton ($\pp$) Large Hadron 
Collider (LHC) is an order of magnitude larger than at the Tevatron. 
The large data sets of top--antitop quark pairs ($\ttbar$) 
that will be collected at LHC, will allow many precision studies. 
The top-quark mass, $\mtop$, is a fundamental parameter of the 
Standard Model (SM) and its precise value is indispensable for 
predictions of cross sections at the LHC. After the Higgs boson 
discovery at the LHC~\cite{Aad:2012tfa,Chatrchyan:2012ufa} and in the 
current absence of direct evidence for new physics beyond the SM, 
precision theory predictions confronted with precision measurements 
are becoming an important area of research for self-consistency tests 
of the SM and in searching for new physics 
phenomena~\cite{Degrassi:2012ry,Baak:2014ora,Alekhin:2012py}.
Due to the high mass the top-quark's width is so large that it 
typically decays before it hadronizes. The $\mtop$ measurements proceed 
then via kinematic reconstruction of the top-quark's decay products, 
a $W$ boson and a $b$-quark jet, and comparisons to Monte Carlo (MC) 
simulations are done. 
These $\mtop$ measurements are often referred to as MC top-quark mass, 
$\mtop^{\rm MC}$, measurements.
There is no immediate interpretation of the measured $\mtop^{\rm MC}$ in 
terms of a parameter of the SM Lagrangian in a specific renormalisation 
scheme. In many Quantum Chromodynamics (QCD) calculations the top quark 
pole mass $\mtoppole$, corresponding to the definition of the mass of 
a free particle, is used as the conventional scheme choice. 
Present studies estimate that the value of $\mtop^{\rm MC}$ differs from 
the $\mtoppole$ by ${\cal{O}}(1\,\GeV$)~\cite{Moch:2014tta,Moch:2014lka}.
The $\mtoppole$ can be measured from inclusive $\ttbar$ production 
cross section ($\sigttbar$)~\cite{Langenfeld:2009wd}. 
However, this $\mtoppole$ determination is currently less precise 
than the achieved $\mtop^{\rm MC}$ measurements. 
This is due to the weak sensitivity of the inclusive
$\sigttbar$ to the $\mtoppole$, but also to the large uncertainties 
on the factorisation and renormalisation scales, the strong coupling 
constant $\alpha_{s}$, and the proton parton distribution function (PDF).
In the following the latest results on the $\mtop^{\rm MC}$, or 
just $\mtop$, measurements in the di-lepton and in the all-jets $\ttbar$ 
decay channel with the ATLAS detector~\cite{atlas} using data at 
$\sqrt{s} = 8\,\TeV$ are presented. 
The data set corresponds to an integrated luminosity of $20.2\,\fbm1$.
The $\mtop$ measurements in di-leptonic $\ttbar$ decay channel, 
where each of the top quarks decays into a $b$-quark, a 
charged lepton and its neutrino, 
is further described in Section~\ref{sec:dileptonic}. 
The $\mtop$ measurement in the all-jets decay channel involves six 
jets, two originating from $b$-quarks and four originating from the 
two $W$ boson hadronic decays.
This recent measurement is detailed in Section~\ref{sec:alljets}.
Finally, the $\mtoppole$ value is determined from inclusive $\sigttbar$ 
measurements in the di-lepton $\ttbar$ decay channel.
This analysis uses data collected at $\sqrt{s} = 7\,\TeV$ and $8\,\TeV$
and corresponding to an integrated luminosity of $4.6\,\fbm1$ and $20.2\,\fbm1$.
The achieved $\mtoppole$ results are summarised in Section~\ref{sec:polemass}.

\section{Di-lepton $\mtop$ measurements at $\sqrt{s}= 8\,\GeV$}
\label{sec:dileptonic}

\begin{figure}[htp]
  \centering
       {\resizebox{0.36\textwidth}{!}{%
        \includegraphics{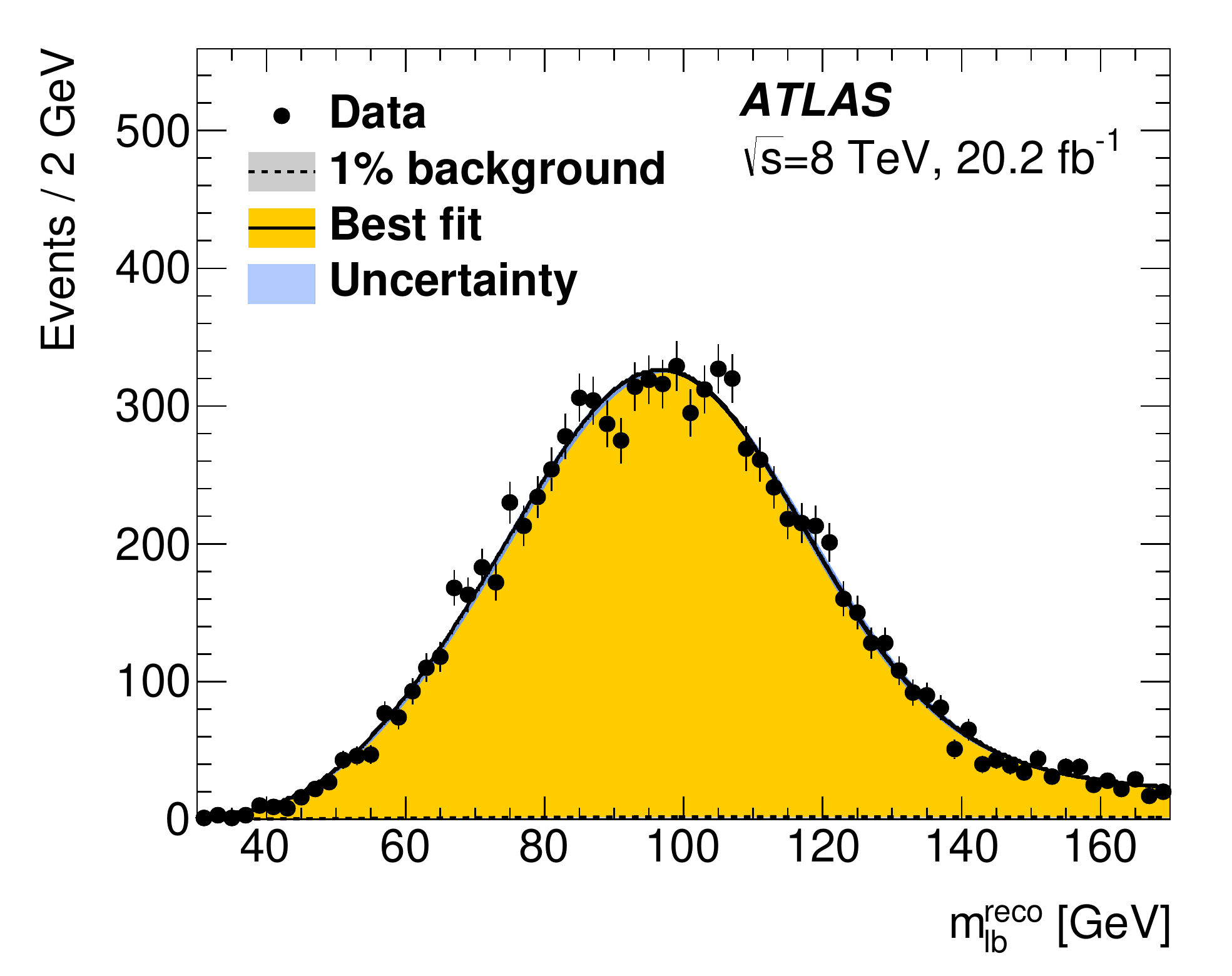}}}
        {\resizebox{0.36\textwidth}{!}{%
        \includegraphics{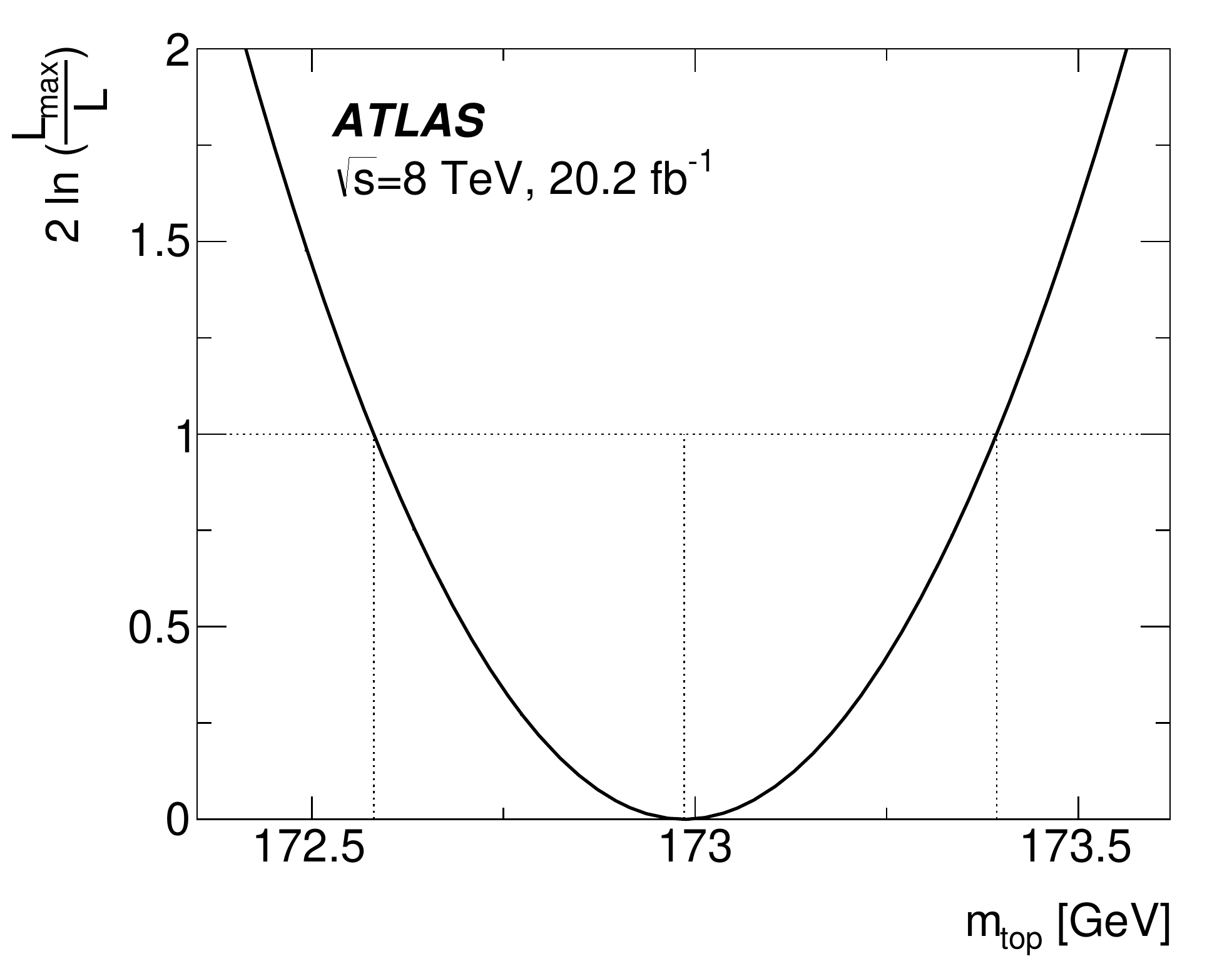}}}
  \caption{
     The left plot shows the distribution for data with statistical 
     uncertainties together with the fitted probability density functions
    for the background alone (barely visible at the bottom of the figure) 
   and for the sum of signal and background. The 
   uncertainty band (invisible) corresponds to the total uncertainty in $\mtop$. 
   The corresponding logarithm of the likelihood as a function of 
   $\mtop$ is displayed in the plot on the right side~\cite{Aaboud:2016igd}.}
  \label{fig:dilepton}
\end{figure}

A new measurement of $\mtop$ is obtained in the $\ttbar\to$
di-lepton decay channel using 2012 data taken at a  
centre of mass energy $\sqrt{s} = 8\, \TeV$~\cite{Aaboud:2016igd}. 
The analysis exploits the decay 
$\ttbar\to \,W^{+}W^{−} b\bar{b}\to\ell^{+}\ell^{-}\nu\bar{\nu}b\bar{b}$, 
where both $W$ bosons decay into a charged lepton and its corresponding 
neutrino.  
In the analysis, the $\ttbar$ decay channels $ee$, $e\,\mu$
and $\mu \mu$ (including $\tau\to e$, $\mu$) are combined and 
referred to as the di-lepton channel. 
Single-top-quark events with the same lepton final states are included 
in the signal. Given the larger data sample compared to 
the $\mtop$ measured at $\sqrt{s}= 7\,\TeV$ in ATLAS~\cite{Aad:2015nba}, 
the event selection was optimised to achieve the smallest total uncertainty.  
The selection from Ref.~\cite{Aad:2015nba} is applied as a pre-selection. 
Here events are required to have a signal from the single-electron or 
single-muon trigger. 
Exactly two oppositely charged leptons are required. 
In the same-lepton-flavour channels, $ee$ and 
$\mu\mu$, a missing transverse momentum, $\met$, $> 60\,\GeV$ is required. 
In  addition, the invariant mass of the lepton pair must satisfy 
 $m^{\ell\ell} > 15 \,\GeV$, and must not be compatible with 
the $Z$ mass within $10\,\GeV$.
In the $e\mu$ channel the scalar sum of the transverse momentum, $\pt$, 
of the two selected leptons and all jets is required to be larger than 
$130\,\GeV$.
The presence of at least two central jets with $\pt > 25\,\GeV$ 
and $|\eta|<2.5$ is required. 
Two $b$-jets taken as originating from the decays of the two 
top quarks are then selected, and two leptons are taken as the leptons from the 
leptonic $W$ decays. From the two possible assignments of the two 
pairs, the combination leading to the lowest average invariant mass 
of the two lepton-–$b$-jet pairs ($\mlb$) is retained.
Starting from this pre-selection, an optimisation of the total uncertainty 
in $\mtop$ is performed.  A phase-space restriction based on the average 
$\pt$ of the two lepton–-$b$-jet pairs ($p_{{\rm T}\ell b}$) is used to obtain the 
smallest total uncertainty in $\mtop$, this corresponds to a cut on 
$p_{T\ell b} > 120\,\GeV$.
To perform the template parameterisation described in Ref.~\cite{Aad:2015nba}, 
an additional selection criterion is applied. The reconstructed $\mlb$
 value is restricted to the range $30\,\GeV < \mlb < 170\, \GeV$.
Using this selection the kinematic distributions in the data are well 
described by the predictions.
The resulting template fit function based on simulated distributions 
of $\mlb$ has $\mtop$ as the only free parameter and an unbinned 
likelihood maximisation gives the $\mtop$ value that best describes 
the data. Figure~\ref{fig:dilepton}, left plot, shows the distribution 
obtained with data together with the fitted probability density 
functions for the background alone that is hardly visible at the 
bottom of the figure. The plot on the right side of 
Figure~\ref{fig:dilepton} shows the final corresponding logarithm of the 
performed likelihood as a function $\mtop$.
This measurement gives 
$\mtop = 172.99\,\pm\,0.41\, (\stat)\,\pm\,0.72\,(\syst)\,\GeV$, 
with the biggest systematic uncertainties coming from jet
energy scale (JES) and relative $b$-to-light-jet energy scale.
This $\mtop$ result is $\sim 40\%$ more precise that $\mtop$ measured
at $\sqrt{s}= 7\,\TeV$.
It is the most precise single result in this decay channel to date.
A combination with the ATLAS measurements in the
$\ttbar\to$ lepton+jets and $\ttbar\to$ di-lepton decay channels from
$\sqrt{s}= 7\,\TeV$ data is performed.
Using a dedicated mapping of uncertainty categories, the combination of
the three measurements results in a value of
$\mtop = 172.84 \, \pm\,0.34 \,(\stat)\,0.61\,(\syst)\,\GeV$.
This result is mostly limited by the calibration of the JES 
and by the Monte Carlo modeling of signal events.

\section{All-jet $\mtop$ measurements at $\sqrt{s}= 8\,\GeV$}
\label{sec:alljets}

\begin{figure}[htp]
  \centering
       {\resizebox{0.36\textwidth}{!}{%
        \includegraphics{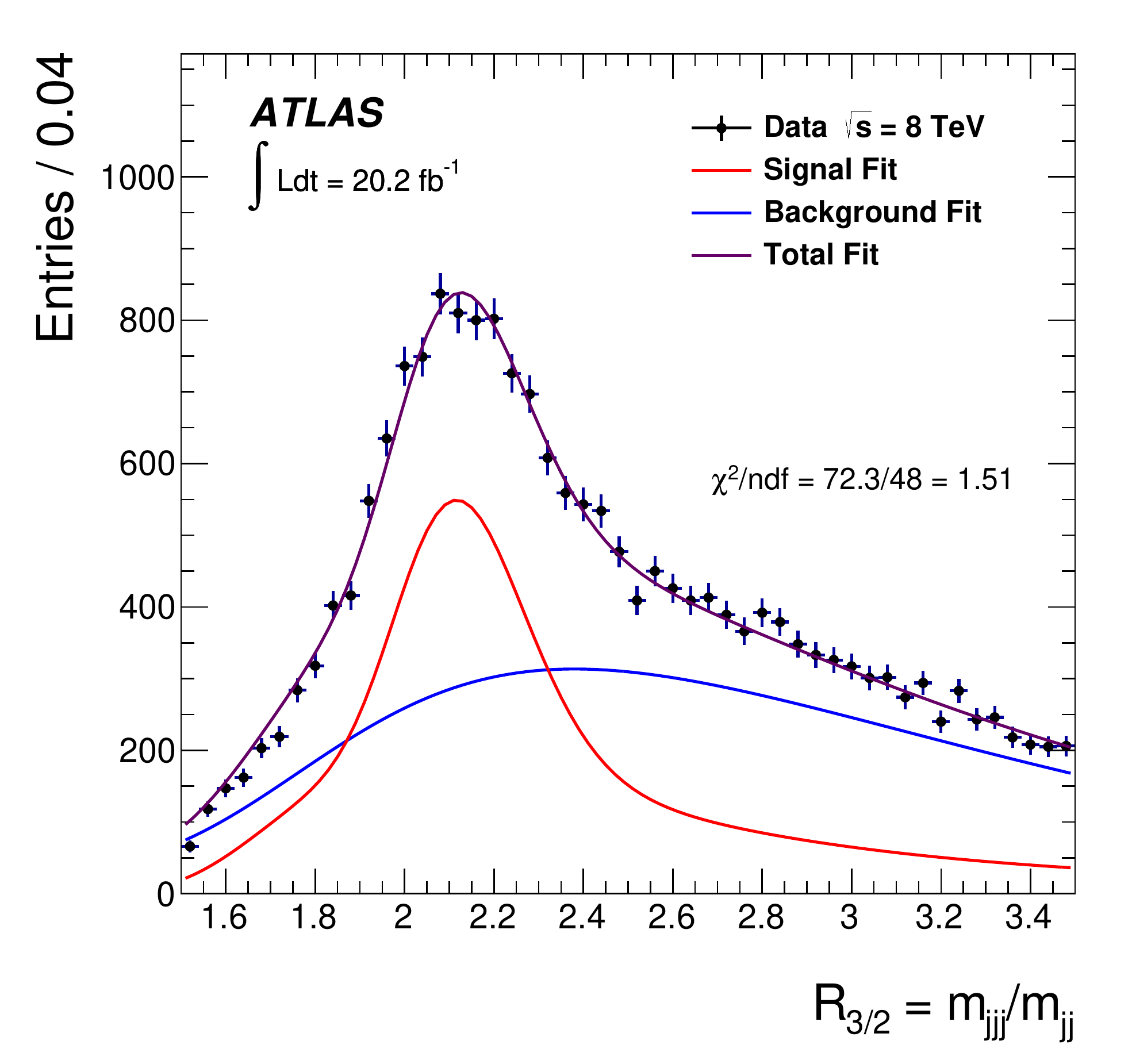}}}
        {\resizebox{0.36\textwidth}{!}{%
        \includegraphics{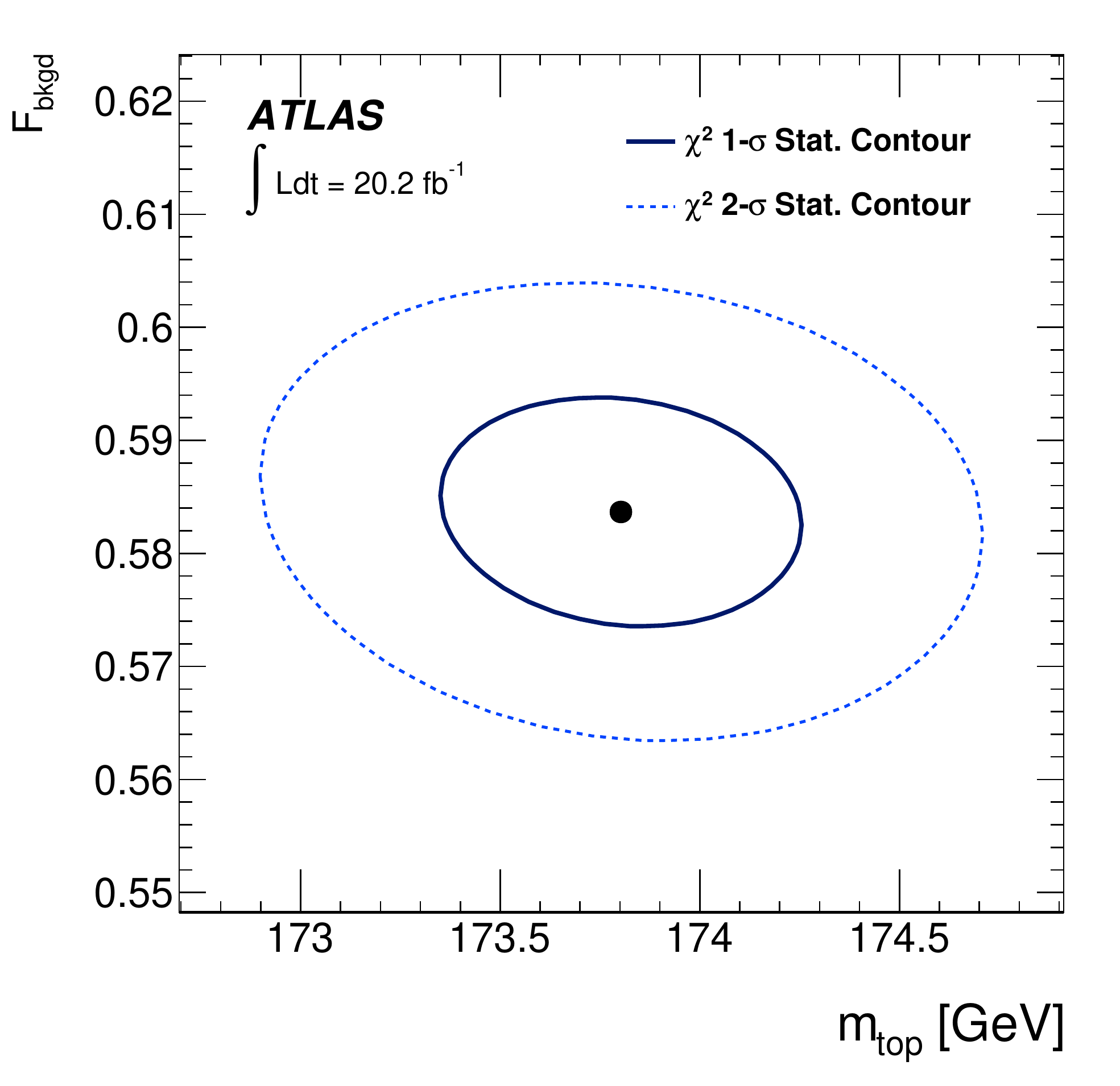}}}
  \caption{
     The left plot shows the $\rtt$ distribution in data with the
     total fit (in magenta) and its decomposition into signal (in red)
     and the multi-jet background (in blue). The errors shown are
     statistical only.
     The right plot shows the ellipses corresponding to the
     $1$-$\sigma$ (solid line) and $2$-$\sigma$ (dashed line)
     statistical uncertainty.
     The central point in the figure indicates the values obtained
     for $\mtop$ on the $x$--axis, and the fitted background fraction,
     obtained within the fit range of the $\rtt$ distribution
     on the $y$--axis.
     The plots do not take into account the small bias correction
     described in Ref.~\cite{Aaboud:2017mae}.
     The final top-quark mass
     is $173.72\,\pm\,0.55\,(\stat)\,\pm\, 1.01\,(\syst)\,\GeV$.}
  \label{alljets}
\end{figure}

A recent $\mtop$ measurement obtained using ATLAS data 
taken at $\sqrt{s} = 8\,\TeV$~\cite{Aaboud:2017mae} exploits the decay
$\ttbar\to\,W^+ b W^- \bar{b}\to q\bar{q}'b q''\bar{q}'''\bar{b}$,
where both $W$ bosons decay into jets from charged 
quarks, $q$.  
This is a challenging measurement to make because of the large multi-jet 
background arising from various other processes of the strong interaction
described by the QCD.
However, all-jets $\ttbar$ events profit from having no neutrinos 
among the decay products, so that all four-momenta can be measured 
directly.
The multi-jet background for the all-hadronic $\ttbar$ 
channel, while large, leads to different systematic uncertainties 
than in the case of the single- and di-leptonic $\ttbar$ channels. 
Thus, all-jets analyses offer an opportunity to cross-check 
$\mtop$ measurements performed in the other channels.
Events in this analysis are selected by a trigger that 
requires at least five jets with $\pt > 55\,\GeV$. 
Events with isolated electrons (muons) with 
$\Et > 25\,\GeV$ ($\pt> 20\,\GeV$) and reconstructed in the 
central region of the detector (within $|\eta| < 2.5$) are rejected. 
To ensure that the selected events are in the plateau 
region of the trigger efficiency curve where the trigger 
efficiency in data is greater than $90$\%, at least five 
of the reconstructed central jets are required to have $\pt > 60\,\GeV$. 
Any additional central jet is required to have $\pt > 25\,\GeV$.
Events containing neutrinos are removed by requiring 
$\met{} < 60\,\GeV$.
In the final selection, events are kept if at least two of the 
six leading transverse momentum jets are identified as originating 
from a $b$-quark ($N_{b_{\rm tag}}$). Such jets are said to be $b$-tagged.
In each event the two jets with leading $b$-tag weights 
($b_i$ and $b_j$) are required to be within an azimuthal angle 
$\Delta \phi(b_i,b_j) > 1.5$.
Finally, another cut based on the azimuthal angle between 
$b$-jets and their associated $W$ boson candidate is applied: the average of 
the two angular separations for each event is required to satisfy 
$\langle\Delta\phi(b,W)\rangle < 2$.
 To determine the $\mtop$ in each $\ttbar$ event, a minimum-$\chi^2$ 
approach is adopted where all possible permutations of the six or 
more reconstructed jets in each event are considered. The permutation 
resulting in the lowest $\chi^2$ value is kept. 
To reduce the multi-jet background in the analysis and to
eliminate events where the top quarks and the $W$
bosons in an event are not reconstructed correctly, a $\chi^2 < 11$ 
is required.
The dominant multi-jet background in the analysis is determined 
directly from the data.
Two uncorrelated variables, the $N_{b_{\rm tag}}$ and the 
$\langle\Delta\phi(b,W)\rangle$, are used to divide the data
events into four different regions, such that the background
is determined in the control regions and extrapolated to the
signal region. The four regions are labeled ABCD and 
distributions of the ratio of three-jet to dijet masses 
($\rtt = m_{jjj}/m_{jj}$) are studied for each of the defined
regions. The $\rtt$ observable is chosen due to its reduced
dependence on the JES uncertainty.
To extract a measurement of the $\mtop$, a template method 
with a binned minimum-$\chi^2$ approach is employed.
For each $\ttbar$ event, two $\rtt$ values are obtained,
one for each $\mtop$ measurement.
Signal and background templates binned in $\rtt$ are
created using simulated $\ttbar$ events, and the data-driven 
background distribution.
After applying a final $\chi^2$ fit, which uses matrix algebra to
include non-diagonal covariance matrices, $\mtop$
is measured to be: 
$\mtop = 173.72\,\pm\,0.55\,(\stat)\,\pm \,1.01\,(\syst)\,\GeV$.
Figure~\ref{alljets} shows the $\rtt$ distribution,
left plot, with the corresponding total fit as well as its
decomposition into signal and the multi-jet background.
The right plot in this figure shows the ellipses corresponding
to $1$-$\sigma$ (solid line) and $2$-$\sigma$ (dashed line)
variations in statistical uncertainty.
The dominant sources of systematic uncertainty in this $\mtop$
measurement, despite the usage of the $\rtt$ observable, come from 
the JES, hadronisation modelling and the
$b$-jet energy scale. 
This measurement is about 40\% more precise than the previous
$\mtop$ measurement performed by ATLAS in the all-hadronic channel 
$\sqrt{s} = 7\, \TeV$~\cite{Aad:2014zea}.

\section{Measurement of $\mtoppole$ in di-leptonic events at $\sqrt{s} = 7$ and $8\,\TeV$}
\label{sec:polemass}

\begin{figure}[htp]
  \centering
       {\resizebox{0.43\textwidth}{!}{%
        \includegraphics{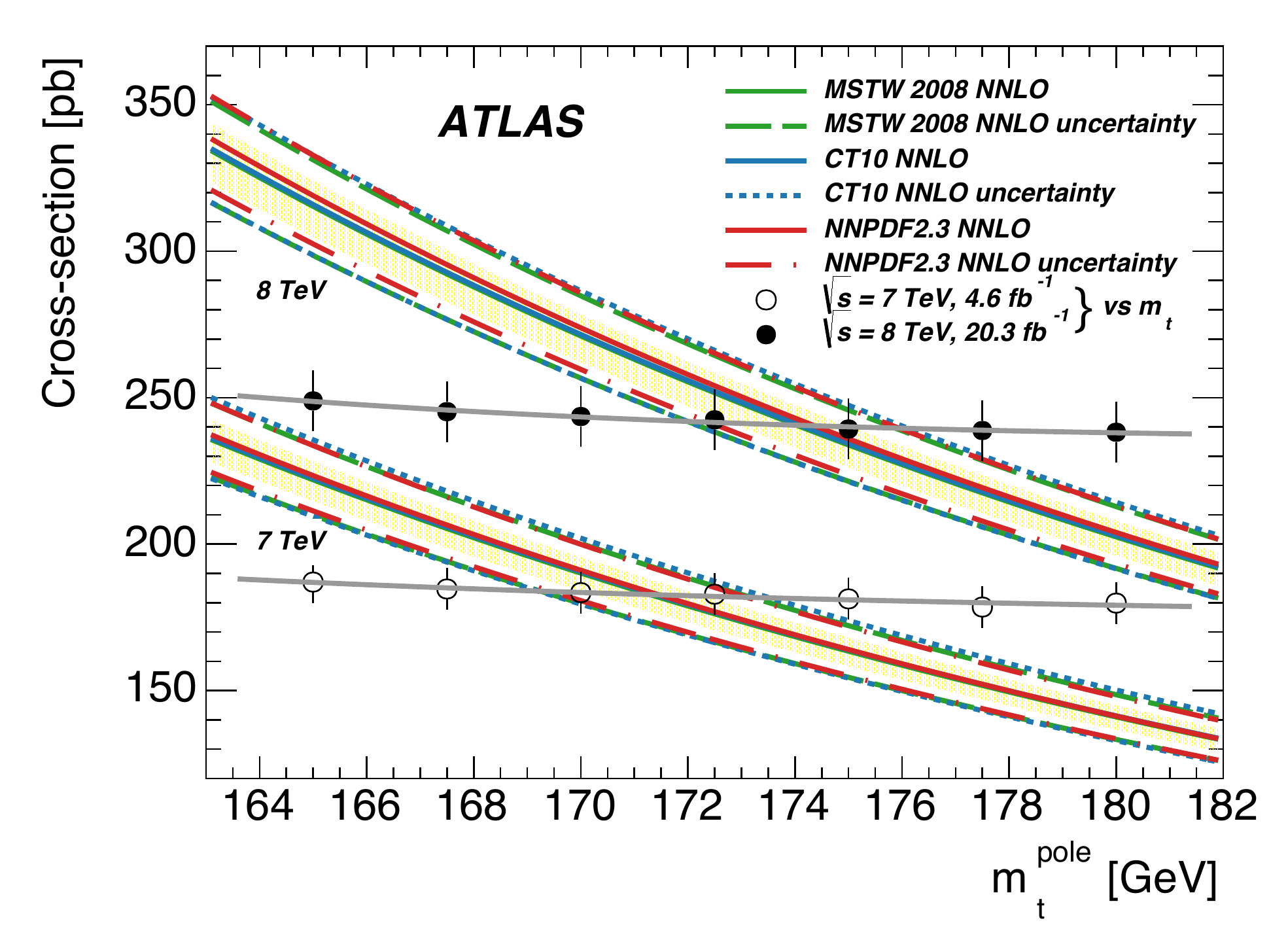}}}
        {\resizebox{0.46\textwidth}{!}{%
        \includegraphics{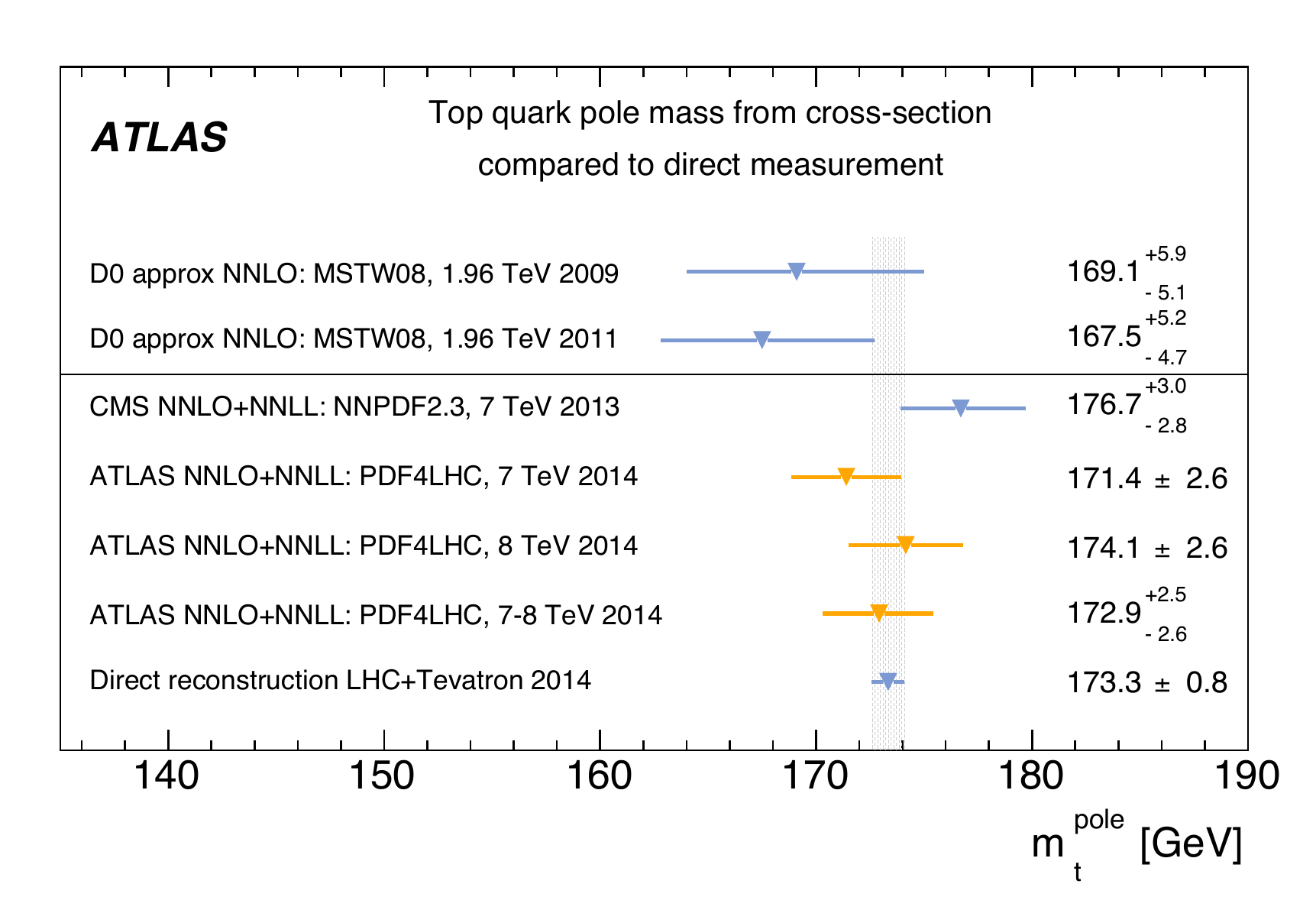}}}
  \caption{Left plot shows predicted NNLO+NNLL production $\sigttbar$
       at $\sqrt{s} = 7\,\TeV$ and $\sqrt{s} = 8\,\TeV$ as a function 
       of $\mtoppole$ showing the central values (solid lines) and total 
       uncertainties (dashed lines) with several PDF sets. 
       The yellow band shows the QCD scale uncertainty. The measurements of 
       $\sigttbar$ are also shown. The right plot shows comparison of 
       $\mtoppole$ values determined from this and previous $\sigttbar$ 
       measurements~\cite{Abazov:2011pta,Chatrchyan:2013haa}, also shown is 
       the $\mtop$ from direct measurements obtained from the 
       LHC+Tevatron combinations~\cite{ATLAS:2014wva}.}
\label{polemass}
\end{figure}

At the LHC, precise measurements of $\sigttbar$ are sensitive to the
the uncertainty on $\alpha_{s}$, to the
gluon parton distribution function (PDF), the $\mtop$, and potential 
enhancements of the cross-section due to physics beyond the Standard Model.
In the following the $\mtoppole$ determination from the inclusive $\sigttbar$ 
measurement in the di-leptonic $e\mu$ channel, 
$\ttbar\to \,W^{+}W^{−} b\bar{b}\to e^{\pm}\mu^{\pm}\nu\bar{\nu}b\bar{b}$, 
is presented~\cite{Aad:2014kva}.
The main background comes from the associated production of a $W$ boson 
and a single top quark, the so called $Wt$ single top background.
The analysis is performed on the ATLAS 2011 - 2012 $\pp$ collision 
data sample, corresponding to integrated luminosities of $4.6\,\fbm1$ 
at $\sqrt{s} = 7\,\TeV$ and $20.3\,\fbm1$ at $\sqrt{s} = 8\,\TeV$. 
Events were required to pass either a single-electron or 
single-muon trigger, with thresholds chosen such that the efficiency 
plateau is reached for leptons with $\pt > 25\,\GeV$.
MC simulated event samples were used to develop the analysis, 
to compare to the data and to evaluate signal and background efficiencies 
and uncertainties.
The analysis makes use of reconstructed electrons, muons and $b$-tagged 
jets. 
A preselection requiring exactly one electron and one muon
was applied. Events with an opposite sign $e \mu$ pair constituted 
the main analysis sample, whilst events with a same-sign $e \mu$ 
pair were used in the estimation of the background from misidentified 
leptons. The production cross-section $\sigttbar$ was determined by
counting the numbers of opposite-sign $e \mu$ events with exactly
one and exactly two $b$-tagged jets and was measured to be 
$\sigttbar = 182.9\,\pm\,3.1\,\pm\,4.2\,\pm\,3.6\,\pm\,3.3\,\pb$ at 
$\sqrt{s} = 7\,\TeV$, and 
 $\sigttbar = 242.4\,\pm\,1.7\,\pm\,5.5\,\pm \,7.5\,\pm\,4.2\,\pb$ at 
$\sqrt{s} = 8\,\TeV$.
where the four uncertainties arise from data statistics, experimental
and theoretical systematic effects related to the analysis, 
knowledge of the integrated luminosity and of the LHC beam energy. 
The strong dependence of the theoretical prediction for $\sigttbar$ 
on $\mtop$, offers the possibility of interpreting measurements of 
$\sigttbar$ as measurements of $\mtop$. 
The theoretical calculations use $\mtoppole$ for predictions. 
The dependence of the cross-section predictions on $\mtoppole$  
is shown in Figure~\ref{polemass}, left plot, at $\sqrt{s} = 7$ and $8\,\TeV$.
The function proposed in Ref.~\cite{Czakon:2013goa} was used to parameterise 
the dependence of $\sigttbar$ on $\mtop$ 
separately for each of the NNLO PDF sets CT10~\cite{Lai:2010vv,Gao:2013xoa}, 
MSTW~\cite{Martin:2009iq} and NNPDF2.3~\cite{Ball:2012cx}, together with 
their uncertainty.
The left plot in Figure~\ref{polemass} also shows the small dependence of the 
experimental measurement of $\sigttbar$ on the assumed value of $\mtop$,
arising from variations in the acceptance and $Wt$ single
top background. A comparison of the theoretical and experimental curves 
shown in this plot allows an unambiguous extraction of $\mtoppole$.
The extraction is performed by maximising using a Bayesian 
likelihood as a function of $\mtoppole$~\cite{Aad:2014kva}.
The likelihood fit maximised separately for each PDF set and centre-of-mass 
energy to give $\mtoppole$ values shown in Table~\ref{tab6}. 
\begin{table}{}
\begin{center}
\begin{tabular}{l||c|c}
\cline{2-3}
\multicolumn{1}{c}{} & \multicolumn{2}{c}{$\mtoppole\,\GeV$ from $\sigttbar$} \\
\cline{2-3}
PDF & \multicolumn{1}{ c }{$\sqrt{s} = 7\,\TeV$ } & \multicolumn{1}{c}{$\sqrt{s} = 8\,\TeV$}\\ 
\hline 
\hline
CT10 NNLO~\cite{Lai:2010vv,Gao:2013xoa}& $171.4\, \pm\, 2.6$  & $171.1\, \pm\, 2.6$\\
MSTW $68\%$ NNLO~\cite{Martin:2009iq}  & $171.2\, \pm\, 2.4$  & $174.0\, \pm\, 2.5$\\
NNPDF2.3 5f FFN~\cite{Ball:2012cx}     & $171.3^{\pm\, 2.2}_{\pm\,2.3}$ & $174.2 \,\pm\,2.4$\\
\hline 
\hline
\end{tabular}
  \caption{Measurements of the $\mtoppole$ determined from 
  the $\ttbar$ cross-section measurements at 
   $\sqrt{s} = 7\,\TeV$ and $\sqrt{s} = 8\,\TeV$ 
  using various PDF sets.}
 \label{tab6}
  \end{center}
\end{table}
A single $\mtoppole$ value was derived for each centre-of-mass 
energy giving
$\mtoppole =171.4\,\pm 2.6\,\GeV$ ($\sqrt{s} = 7\,\TeV$) and 
$\mtoppole = 174.1\,\pm 2.6\,\GeV$ ($\sqrt{s} = 8\,\TeV$).
Considering only uncorrelated experimental uncertainties,
the two values are consistent at the level of $1.7$ standard
deviations.
Finally, $\mtoppole$ was extracted from the combined
$\sqrt{s} = 7\,\TeV$ and $\sqrt{s} = 8\,\TeV$ dataset using 
the product of likelihoods for each centre-of-mass 
energy and accounting for correlations via nuisance parameters.
The resulting value using the envelope of all three considered PDF 
sets is $\mtoppole = 172.9^{+2.5}_{-2.6}\,\GeV$.
All extracted values are consistent with the $\mtop$ or $\mtop^{\rm MC}$ 
measurements obtained from kinematic reconstruction of $\ttbar$ events, 
see right plot in Figure~\ref{polemass}.


%
%
%


\end{document}